\pdfoutput=1
\documentclass[11pt,a4paper]{article}
\usepackage{jheppub}
\usepackage[british]{babel}
\usepackage[latin1]{inputenc}
\usepackage[T1]{fontenc}
\usepackage[final]{showkeys} % [draft/final]
\usepackage{enumerate}
\hypersetup{pdfauthor={M. Cadoni, A. M. Frassino and M. Tuveri},
            pdftitle={On the universality of  thermodynamics and $\eta/s$ ratio for  the charged  Lovelock black brane}
            }
\usepackage[]{cleveref}
\Crefname{equation}{Eq.}{Eqs.}
\crefformat{plural}{#2eqs.~(#1)#3}
\makeatletter\g@addto@macro\bfseries{\boldmath}\makeatother%

\newcommand{\be}{\begin{equation}}
\newcommand{\ee}{\end{equation}}
\newcommand{\bea}{\begin{eqnarray}}
\newcommand{\eea}{\end{eqnarray}}

\def\a{\alpha}
\def\s{\sigma}

\def\oc{\omega_5}
\def\lb{\label}

\begin{document}

% title page
%\preprint{\today}

\title{ On the universality of  thermodynamics and $\eta/s$ ratio for  the charged  Lovelock black branes}

\author[1,2]{Mariano Cadoni,}\emailAdd{mariano.cadoni@ca.infn.it}
\author[3,4]{Antonia M. Frassino}\emailAdd{frassino@fias.uni-frankfurt.de}
\author[1,2]{and Matteo Tuveri}\emailAdd{matteo.tuveri@ca.infn.it}

\affiliation[1]{Dipartimento di Fisica, Universit\`a di Cagliari,\\Cittadella Universitaria, 09042 Monserrato, Italy}
\affiliation[2]{INFN, Sezione di Cagliari}
\affiliation[3]{Frankfurt Institute for Advanced Studies,\\ Ruth-Moufang-Str. 1, D-60438 Frankfurt am Main, Germany}
\affiliation[4]{Institut  f\"{u}r Theoretische Physik, Johann Wolfgang
Goethe-Universit\"{a}t,\\ Max-von-Laue-Stra\ss e 1, D-60438 Frankfurt am Main, Germany}

%\keywords{keyword one, keyword two}
\arxivnumber{}

\abstract{
We investigate general features of charged Lovelock black branes by giving a detailed description of geometrical, thermodynamic and 
holographic properties of charged Gauss-Bonnet (GB) black branes in five dimensions. 
We show that when expressed in terms of effective physical parameters, the thermodynamic behaviour 
of charged GB black branes is completely indistinguishable from that of charged Einstein black branes. 
Moreover, the extremal, near-horizon limit  of the two classes of branes is exactly the same as  
they allow for the {\sl same} AdS$_2\times R_3$, near-horizon, {\sl exact} solution.
This implies that, although in the UV the associated dual QFTs are different, 
they flow in the IR to the same  fixed point.  The calculation of the shear viscosity to entropy ratio $\eta/s$
 confirms these results.
Despite the GB dual plasma  has in general a non-universal temperature-dependent $\eta/s$, it flows monotonically 
to the universal value $1/4\pi$ in the IR.  
For  negative (positive) GB coupling constant,  $\eta/s$ is an increasing (decreasing) function of the  
temperature and the flow respects (violates)  the KSS bound.
} 

\maketitle

%\date{\today}

\section{Introduction}

Since its  first formulation, Lovelock gravity \cite{Lovelock01}   has been a fruitful and widely explored subject
\cite{ Cai:2003kt,Castro:2013pqa,Camanho:2011rj,Takahashi:2011du}.
The peculiarity of the theory is to be a higher curvature gravity theory with second-order field  equations for the metric.
This nice feature not only allows to avoid some  of the shortcomings of generic  higher-derivative theories 
(such as ghosts in  the linearized excitation spectrum and  ill-posed Cauchy problem) but also enables us to derive exact  
black hole (and black brane) solutions of the theory.   
As a consequence,  the thermodynamics of Lovelock black holes  is well known and has several interesting, nontrivial features. 
One of these features
is that the thermal entropy \cite{ Myers:1988ze, Wald1994} and the holographic entanglement entropy \cite{Caceres:2015bkr}  of a black hole depend on the higher-curvature gravitational couplings. It is also well understood that there are in these theories new types of phase transitions that also depend on 
the value of the gravitational couplings  \cite{Camanho:2013uda, Frassino:2014pha, Dolan:2014vba,Cai:2001dz}.

Lovelock gravity is interesting also from the holographic point of view. The higher curvature terms in the action correspond, on the gauge theory side of the AdS/CFT correspondence,  
to corrections due to finite $\cal{N}$ 
(rank of the gauge group) and finite t'Hooft coupling $\lambda_{tH}$. 
Thus, Lovelock gravity allows  to investigate finite $\cal{N}$ and $\lambda_{tH}$ effects without having some of the undesirable 
features of higher curvature gravity theories.

Among the Lovelock gravity theories, one of the most investigated cases, that will also be the subject of this paper, is the five-dimensional (5$d$)  Gauss-Bonnet (GB) theory.
Specifically, GB gravity is the 2nd-order Lovelock gravity, \textit{i.e.}, it includes only quadratic curvature corrections in the Einstein-Hilbert action.
The main reason to study 5$d$ GB in the AdS/CFT framework is that the dual QFT lives in four spacetime dimensions. Hence,  5$d$ GB gravity can be used  to describe  $1/\cal{N}$ corrections to relativistic QFTs with a gravitational dual.  
Particular attention has been devoted to the low-frequency hydrodynamic limit, $\omega, k<<T$, where 
$\omega$ is the frequency, $k$ is the wavelength number and $T$ is the temperature of the dual thermal QFT.
In this limit, the theory describes a sort of ``GB plasma''  for which  transport coefficients can be calculated using the rules of 
the AdS/CFT correspondence.   

A quantity, which plays  a distinguished role in the hydrodynamic regime of thermal QFTs with gravitational duals  is the shear viscosity to entropy density ratio $\eta/s$. It has been shown that $\eta/s$ attains  an \textit{universal} value $1/4\pi$  for all gauge theories with Einstein gravity duals 
\cite{Policastro:2001, Edalati:2009bi, Buchel:2004qq, Benincasa:2006fu, Kats:2007mq, Landsteiner:2007bd,Iqbal:2008by, Buchbinder:2008nf}. 
This fact  motivated the formulation of a fundamental bound  $\eta/s\ge 1/4\pi$, known as Kovton, Son and Starinets (KSS) bound \cite{KSS:2003, KSS:2005}, which also found support from energy-time uncertainty principle arguments in the weakly coupled regime \cite{KSS:2005} and known experimental data for quark-gluon plasma \cite{Song:2010mg,KSS:2005}.
However, it was  soon realized that higher curvature gravity theories may generically violate the bound \cite{Brigante-Myers:2008}. 
This is, in particular, true for GB gravity theories with a positive  coupling constant. 

Violation of the KSS bound of higher curvature gravity theories can be understood as generated by finite-$\cal{N}$, 
finite-$\lambda_{tH}$ effects and traced back to the inequality of the two central charges of the dual QFT {\cite{Buchel:2008vz,Buchel:2004di}. 
Nevertheless, this does not answer the question about the possible existence of general bounds on $\eta/s$ lower than the KSS one.
The GB gravity, owing to its features, is the most promising playground for trying to answer this  question. 
Progress in this direction has been achieved  by imposing causality and positivity  of energy to the QFT dual to 
GB gravity \cite{Buchel:2009tt,Brigante:2008gz,Hofman:2009ug}.
These requirements imply  some constraints on the  GB coupling parameter, which in turn translate into a bound on $\eta/s$ 
lower than the KSS bound \cite{Buchel:2009tt,Brigante:2008gz,Hofman:2009ug}. However, the hydrodynamic  transport coefficients   
of a theory  are expected to be 
 determined by  
IR physics  whereas causality requirements are in the domain of  the UV behavior of the dual QFT. The existence of a  
fundamental bound of the previous kind  for the GB  plasma  would, therefore, imply an  interplay between IR and UV physics, whose meaning is  presently not clear.

In a parallel, very recent,  development it has been shown that generically the KSS bound is violated if translation invariance is broken \cite{Hartnoll:2016tri, Davison:2015taa,Rebhan:2011vd,Mamo:2012sy}. 
If translation symmetry is preserved in the IR, $\eta/s$  tends to a constant  as $T\to0$, whereas it scales as a positive power of $T$ when translation symmetry is broken. 
Although the breaking of translation  symmetry prevents a purely hydrodynamic interpretation of $\eta$, this result  strongly indicates that  bounds on $\eta/s$ are  completely determined by IR physics and insensitive to the UV regime of the theory. 

A  promising way to tackling this kind of problems is to consider gravitational backgrounds in  which $\eta/s$ flows as a function of the temperature and for which an IR fixed point exists at $T=0$. Following this indication, in this paper, we will focus on the charged 5$d$ GB black brane solutions (BB) for which it is known that the ratio $\eta/s$ flows as a function of the temperature \cite{Cai:2008ph}. 

We will start by investigating the general Lovelock BB solution as a thermodynamic system. We will show  that, when expressed in terms of effective physical parameters, the thermodynamic behavior of charged Lovelock BB is completely indistinguishable from that of charged Einstein BB.
We then proceed by focusing on the 5$d$ GB case and investigating in detail the geometrical properties of the charged GB black brane. We show that the theory allows for two branches of solutions continuously connected trough a branch-point singularity. Holographically they represent flows between  two different CFTs through a  singularity.
Moreover, we  show that  at extremality, in the near horizon regime,  
the  charged GB black brane has \textit{exactly the same  AdS$_2\times R_3$  geometry} of the Einstein charged black brane.
In fact, in the near horizon regime the contributions of the higher-curvature terms  to the field equations vanish and the AdS$_2\times R_3$  solution of Einstein-Maxwell  
gravity in 5$d$ is also the exact solution of GB gravity in 5$d$.

In terms of the dual QFT description this  means that, although in the UV the associated dual QFTs for Einstein and GB gravity are different, in the IR they flow to the same  fixed point. 
We then calculate  the shear viscosity to entropy ratio $\eta/s$ for the extremal and non-extremal case, using the simple method recently proposed in Refs. \cite{Lucas:2015vna,Hartnoll:2016tri, Davison:2015taa}.
Whereas in the  non-extremal case we find a non-universal, monotonically increasing (for negative GB coupling parameter) or decreasing (for positive GB coupling parameter) temperature-dependent expression for  $\eta/s$, in the extremal case we find  the universal value $1/4\pi$. Thus, charged Gauss-Bonnet gives an example of a higher curvature gravity theory  in which the IR behaviour of the dual theory respects the universal bound for $\eta/s$ and is completely independent from the  UV regime. 
 
The structure of the paper is as follows. In Sect. \ref{sec:setup} we briefly review some relevant features of  black brane solutions of Lovelock-Maxwell gravity and show the universality of their thermodynamic behaviour. In Sect. \ref{sec:RN} we review the Reissner-Nordstr\"{o}m (RN) BB solutions of 5$d$ Einstein-Maxwell, including its extremal limit and its AdS$_2\times R_3$, extremal, near horizon geometry. In Sect. \ref{sec:Brane5d} we discuss the  charged black brane solution  of GB gravity, paying particular attention to the geometry of the solution and the extremal, near horizon regime. In  Sect. \ref{thermodynamics} we discuss the charged GB black brane thermodynamics, and we consider in detail the thermodynamic behaviour at small and large temperature. In Sect. \ref{General_eta/s} we discuss the shear viscosity to entropy ratio for the GB plasma and compute the value both for $T\neq 0$  and $T=0$. 
We also present a discussion about the large $T$ and small $T$ behaviour.
Finally, in Sect. \ref{sec:conclusions} we draw our conclusions. 
In the Appendix  \ref{sec:BH} we briefly discuss the black hole solutions  of the GB theory, \textit{i.e.}, the solution with spherical horizons.    

\section{Black brane solutions of Lovelock gravity} \label{sec:setup}

Let us consider black branes  that are solutions of Lovelock higher curvature gravity in $d$-dimensional spacetime. 
To describe the static, electrically charged, radially symmetric AdS Lovelock BB, we use the following line element and electromagnetic (EM) field
 \be\label{solution}
ds^{2} = -f\left(r\right)N^2dt^{2}+f\left(r\right)^{-1}dr^{2}
+\frac{r^2}{L^2} d\Sigma_{ d-2}^{2}\,, 
\qquad  F= \frac{Q}{r^{d-2}} dt\wedge dr\,, 
\ee
where $d\Sigma_{d-2}^{2}$   denotes the $\left( d-2 \right)$-dimensional space with zero curvature and planar topology,
whereas $L$ is related to the cosmological constant ${\hat{\alpha}_{(0)}}$ by $L^{-2}= {\hat{\alpha}_{(0)}}/(d-1)(d-2)$. 

Notice that the metric in Eq.\eqref{solution} differs from that in the usual Schwarzschild gauge  by a (constant) rescaling $t\to Nt$  of 
the  time coordinate $t$.
As we will see later in this paper this rescaling is necessary  in order to have a unit speed of light  in the dual CFT. 
Using the rescaled Lovelock coupling constants 
\begin{equation} \label{couplCONST}
L^{-2}=\alpha_{0}=\frac{\hat{\alpha}_{(0)}}{\left(d-1\right)\left(d-2\right)}\,,\quad{\alpha}_{1}={\hat \alpha}_{(1)}\,,\quad\alpha_{k}=\hat \alpha_{(k)}\prod_{n=3}^{2k}\left(d-n\right)  {\quad\mbox{for}\quad  k\geq2}\,,
\end{equation}
the field equations read 
\begin{equation}
\sum_{k=0}^{k_{max}}\hat{\alpha}_{\left(k\right)}\mathcal{G}_{ab}^{\left(k\right)}=
8\pi G_N \Bigl(F_{a c} {F_b{}^{\,c}} -\frac{1}{4}g_{a b} F_{c d}F^{c d}\Bigr)\,.
 \label{eq:Graveq}
\end{equation}
where $G_{N}$ is the $d$-dimensional Newton's constant and each of the Einstein-like tensors 
$\mathcal{G}_{\,\,\quad b}^{\left(k\right)\, a}$ defined by 
\begin{equation}
\mathcal{G}_{\,\,\quad b}^{\left(k\right)\, a}=-\frac{1}{2^{\left(k+1\right)}}\delta_{b\, e_{1}f_{1}\ldots e_{k}f_{k}}^{a\, 
c_{1}d_{1}\ldots c_{k}d_{k}}R_{c_{1}d_{1}}^{\quad e_{1}f_{1}}\ldots R_{c_{k}d_{k}}^{\quad e_{k}f_{k}}\label{eq:G}\,,
\end{equation}
 independently satisfies a conservation law $\nabla_{a} \mathcal{G}_{\,\,\quad b}^{\left(k\right)\, a} =0 $.
The higher-curvature terms contribute to the equations of motion only for $d > 2k$.  For $d = 2k$ the higher-curvature corrections are  topological, 
and they vanish identically in lower dimensions.
%The higher-curvature terms contribute to the equations of motion for $d > 2k$, whereas they are topological in $d = 2k$, 
%and vanish identically in lower dimensions. 
Setting $\hat{\alpha}_{(k)} = 0$ for $k \geq 2$, one can recover the standard form of general relativity. 
%General relativity is recovered upon setting $\hat{\alpha}_{(k)} = 0$ for $k \geq 2$.
%
In the notation \eqref{couplCONST}, the field equations \eqref{eq:Graveq} reduce to the requirement that $f\left(r\right)$ 
solves the following polynomial equation of degree $k_{max}= \left [ \frac{d-1}{2} \right ]$ (see \textit{e.g.}, 
\cite{Cai:2003kt,Castro:2013pqa,Camanho:2011rj,Takahashi:2011du,Boulware:1985wk, wheeler1986symmetric1, wheeler1986symmetric2})

 \begin{equation} \label{eq:poly}
{\cal P}\left(f\right)=\sum_{k=0}^{k_{max}}\alpha_{k}\left(\frac{\kappa-f}{r^{2}}\right)^{k}=
\frac{\omega_{d}M_{ADM}}{Nr^{d-1}}-
\frac{8\pi G_{N}Q^{2}}{(d-2)(d-3)}\frac{1}{r^{2d-4}}\,.
\end{equation}
 Here $M_{ADM}$ is the ADM mass of the black brane and $\omega_{d}$ is
\begin{equation}\label{eq:omega_d}
\omega_{d}=\frac{16\pi G_{N}}{(d-2)}\frac{L^{d-2}}{V^{d-2}}\,
\end{equation}
where 
$V^{d-2}$ is the volume of the $(d-2)$-dimensional space with curvature $\kappa=0$. 
The electric charge $Q$ of the brane is  
\begin{equation}\label{MQ}
Q = \frac{L^{d-2}}{2V_{d-2}} \int * F\,. 
\end{equation}

%----------------------------------------------------------------------------------
\subsection{ Universality  of black brane thermodynamics in Lovelock gravity}
\lb{sect:ub}
Interestingly, even without knowing $f=f(r)$ in Eq.\eqref{eq:poly} explicitly, it is possible to find the thermodynamic quantities characterizing the Lovelock black 
brane solution \cite{Cai:2003kt,Jacobson:1993xs, Kastor:2011qp}. 
Let $r_+$ denotes the radius of the event horizon, determined as the largest root 
of $f \left(r\right)=0$. 
Introducing the effective mass $M$ and temperature $T$ related to the usual ADM mass $M_{ADM}$ and Hawking temperature $T_H$ by the relations 
\be\lb{eff}
M=\frac{M_{ADM}}{N},\quad T= \frac{T_H}{N},
\ee
  the black  brane mass $M$, the temperature $T$, the entropy $S$, and the gauge potential $\Phi$ are given by \cite{Cai:2003kt, Ge-Sin:2009} 
\bea
M&=&\frac{1}{\omega_dL^{2}}r_+^{d-1}+\frac{V_{d-2}}{2(d-3)L^{d-2}}\frac{Q^2}{r_+^{d-3}}\,,\label{eq:Mass}\\
%T &=&  \frac{\vert f^\prime(r_+)\vert}{4\pi} =\frac{1}{4\pi r_+ D(r_+)}\left[\sum_k\kappa\alpha_k(d\!-\!2k\!-\!1)\Bigl(\frac{\kappa}{r_+^2}\Bigr)^{k-1}\!\!\!-\frac{8\pi G_N Q^2}{(d-2)r_+^{2(d-3)}}\right]\,,\qquad  \label{T}\\
T&=& \frac{1}{2\pi N}\frac{1}{\sqrt{g_{rr}}}\frac{d\sqrt{-g_{tt}}}{dr}\Big|_{r=r_+}\nonumber \\
&=&\frac{1}{4\pi r_+}\left[(d-1)\left(\frac{r_+}{L}\right)^2-\frac{8\pi G_N Q^2}{(d-2)r_+^{2(d-3)}}\right]\,,\qquad  \label{T}\\
S&=&\frac{V^{d-2}}{4G_N}\left(\frac{r_+}{L}\right)^{d-2}\,, \quad 
 \Phi=\frac{V_{d-2}}{(d-3)L^{d-2}}\frac{Q}{r_+^{d-3}}\,. \label{eq:Entropy}
\eea
The  rescaling of the physical parameters (\ref{eff}) of the Lovelock BB having the dimensions of energy 
is essentially due to the presence of the constant $N^2$ in the metric. The two time coordinates $t$ and $Nt$ correspond to 
using two different units to measure the energy.
However, when we deal with Einstein-Hilbert branes the rescaling of the time coordinate is not necessary 
and we will simply set $M=M_{ADM}$ and $T= {T_H}$.
Notice that the area-law for  the entropy  $S$ always hold for the generic  Lovelock black brane.

A striking feature of these thermodynamic expressions is that {\sl they do not depend  on the Lovelock coupling constants $\alpha_k$ for $k\ge2$} 
but only on $\alpha_0$ and $\alpha_1$, \textit{i.e.}, they depend only on the cosmological constant and on Newton 
constant. This means that the thermodynamic behaviour of the BB in Lovelock theory is universal, in the sense  that 
{\sl it does not depend on the higher order curvature terms} but only on the Einstein-Hilbert term, the cosmological constant and the matter 
fields content (in our case the EM field). 
This implies, in turn, that as thermodynamic system 
the charged BBs of Lovelock gravity are indistinguishable from the Reissner-Nordstr\"{o}m BBs of Einstein-Hilbert gravity.
Notice that this feature is not shared by the black hole solutions of the theory, \textit{i.e.}, solutions with spherical or hyperbolic horizons.
In fact, in the Lovelock thermodynamic expressions (see Refs. \cite{Cai:2003kt, Ge-Sin:2009}) the  dependence on the Lovelock coupling 
constants $\alpha_{k\ge2}$ is introduced by the dependence on the curvature $\kappa$ of 
the $(d-2)$-dimensional spatial  sections. This dependence drops out when $\kappa=0$. 

We remark, however, that the universal thermodynamic behaviour of charged Lovelock black branes is strictly true only when we 
choose $N=1$ in the metric \eqref{solution}. 
As we will see later in this paper, the parameter  $N$ has to be fixed in terms of  the Lovelock coupling constants 
$\alpha_{k\ge2}$. Hence, the  ADM mass and the Hawking temperature of the Lovelock BB will depend  on $\alpha_{k\ge2}$. 
The universality of the Lovelock BB thermodynamics is recovered simply by rescaling the units we use to measure the energy, 
\textit{i.e.}, by using in Eqs. \eqref{eq:Mass} and \eqref{T} the effective parameters $M$ and $T$ instead of $M_{ADM}$ and $T_H$.  

In the following, we provide a detailed calculation for the case $k_{max}=2$, \textit{i.e.}, GB gravity in five spacetime dimensions, 
which is the most interesting case from the AdS/CFT point of view. 
However, we expect that most of our considerations can be easily generalized to every charged BB solution of Lovelock gravity in generic dimensions.

%--------------------------------------------------------
\section{$5d$ Reissner-Nordstr\"{o}m  black brane solution}
\lb{sec:RN}
Let us preliminary review some known facts about the RN BB solutions of Einstein-Maxwell gravity.
Setting $\a_k=0$ for $k\ge2$ and $d=5$ in Eq \eqref{eq:G}, we have standard GR equations sourced by an electromagnetic field. 
For this choice of the parameters, Eq. (\ref{eq:poly})  is a linear equation in $f$ that gives the following solution:
\be%\lb{rnb}
\label{eq:fRN}
f= \a_0 r^2- \frac{\oc M}{r^2}+\frac{4\pi}{3} \frac{G_NQ^2}{r^4},
\ee
where $\omega_{5}$ is given by Eq. \eqref{eq:omega_d} and $G_{N}$ is the five dimensional Newton's constant.
%\subsection{Asymptotic geometry}
Performing the asymptotic limit  $r \rightarrow \infty$,  the function \eqref{eq:fRN} reduces to $f=r^2/L^2$,
\textit{i.e.},  AdS$_5$  with AdS length  $L^2=\a_0^{-1}$.
The ratio $L^3/G_N$ is proportional to the central charge $c$ of the dual CFT. 
The central charge $c$ can be defined as the coefficient of the large temperature expansion of the  free energy 
(see Sect. \ref{sect:LT}). The condition for the validity of  classical AdS gravity in the bulk is $c>>1$. 
In most of the established examples of the AdS/CFT correspondence  $c\propto {\cal{N}}$, where the limit $c>>1$ is 
referred to as the large $ {\cal{N}}$ limit.
%\subsection{Horizons}

Setting $r^2=Y$ in Eq. \eqref{eq:fRN}, the RN BB horizons are determined by the cubic equation
\be\lb{hh}
Y^3-\oc ML^2 Y+\frac{4\pi}{3}G_N L^2Q^2=0.
\ee
This equation has two positive roots for
\be\lb{bbrn}
 M^3\ge 12\pi^2\frac{G_N^2Q^4}{\oc^3L^2},
\ee
which gives the extremal (BPS  \cite{B, PS}) bound for the RN black brane  in $5d$.
In general, we will have an inner and outer horizon, when the bound is saturated the two horizons merge 
at $r_0$ and the RN BB becomes extremal.
%
%\subsection{Extremal brane  and near-horizon geometry}
In the extremal case,  Eq. \eqref{hh} has a double root at $Y_0=\sqrt{\oc ML^2/3}$ 
and $f\left ( r \right)$ can be factorized in the following way
\be\lb{ll}
f(r)=\frac{1}{L^2 r^4}\left(r^2+r_0^2\right)\left(r-r_0 \right)^2\left(r+r_0 \right)^2,\quad r_0=\left(\frac{\oc ML^2}{3}\right)^{1/4}.
\ee
The extremal near-horizon geometry can be determined expanding the metric near $r_0$ and keeping only the leading term in the metric%. We have
\be\lb{nhrn}
f(r)= \frac{12}{L^2} (r-r_0)^2,
\ee
a simple translation of the radial coordinate $r\to r+r_0$ gives the AdS$_2\times R_3$ extremal near-horizon geometry with 
AdS$_2$ lenght $l$

\be\lb{nhg1}
ds^2= -\left(\frac{r}{l}\right)^2 dt^2+\left(\frac{l}{r}\right)^2 dr^2+ \left(\frac{r_0}{L}\right)^2 d\Sigma^2_3, \, l^2=\frac{L^2}{12}.
\ee
The extremal solution given in Eq. \eqref{ll} is a soliton interpolating between the asymptotic AdS$_5$  geometry in the 
UV and the AdS$_2\times R_3$ geometry \eqref{nhg1} in the IR.

\section{Gauss-Bonnet  solution} \label{sec:Brane5d}
We use the form \eqref{solution} with coupling constant \eqref{couplCONST}.
For $k=2$ and generic curvature $\kappa$, Eq. \eqref{eq:poly} reduces to a quadratic equation
\bea \label{eq:polyGB5D}
\alpha_{2}\frac{(\kappa-f)^{2}}{r^{4}}+\frac{(\kappa-f)}{r^{2}}+\alpha_{0}-\frac{\omega_{d}M}{r^{d-1}}+
\frac{8 \pi G_N Q^{2}}{(d-2)(d-3)r^{2d-4}}=0\,,
\eea
from which one obtains 
two possible solutions, $f_\pm$. In the following, we will refer to the solution $f_-$ as the `{\em Einstein branch}' because it  approaches the Einstein 
case when the Gauss--Bonnet coupling $\alpha_2$ goes to zero and
to $f_+$ as the `{\em Gauss--Bonnet branch}' \cite{Frassino:2014pha}.
The quadratic Eq. \eqref{eq:polyGB5D} gives the following necessary condition requirement for the existence of $f_\pm$ for large $r$: 
\be\lb{v1}
1-4\alpha_0\alpha_2\geq 0\,. 
\ee
When this inequality is violated, the space becomes compact because of the strong nonlinear curvature \cite{Frassino:2014pha}. Therefore, there is no asymptotic `AdS region' and consequently no proper black hole with standard asymptotics.

\subsection{$5d$ GB black brane}
In this subsection, we discuss the special case of $5d$ GB BB ($\kappa=0$). 
Moreover, from now on we set $\alpha_1=1$ in order to recover the usual Newtonian limit.
It is easy to check that that for  $d=5$ and $\kappa=0$, then Eq. \eqref{eq:polyGB5D} reduces to the following equation
\bea
\alpha_{2}\frac{f{}^{2}}{r^{4}}-\frac{f}{r^{2}}+\alpha_{0}-\frac{\omega_{5}M}{r^{4}}+\frac{4\pi Q^{2}}{3r^{6}}=0\,
\eea
and the two branches are respectively
\be\lb{f}
f_{\pm}=\frac{r^2}{2\a_2}\left[1\pm \sqrt{1-4 \a_0\a_2}\sqrt{1+\frac{4M\a_2\omega_{5}}{(
1-4 \a_0\a_2)}\frac{1}{r^4}-\frac{16\pi G_N}{3}\frac{Q^2\a_2}{
1-4 \a_0\a_2}\frac{1}{r^6}}\right].
\ee
%
%\subsection{Asymptotic geometry}
%
In case of positive GB coupling $\alpha_2>0$ that satisfy the condition \eqref{v1}, % ( $0<\a_2 \le 1/(4\a_0)$ ) 
the two branches describe two asymptotically AdS$_5$ spacetimes, 
however, from Eq. \eqref{f} one can see that $f_+$ has no zeroes, hence the $f_+$-branch does not describe a BB but a solution with no 
event horizon.  Thus, only the $f_-$-branch describes a BB solution.\\ 
Let us now study the asymptotic geometry of the GB BB.
At leading order for $r\to \infty$ 
the metric coefficient $g_{tt}= N^2 f \left( r \right)$ in Eq.\eqref{solution} becomes 
\begin{equation}
g_{tt}\to N^2\frac{r^2}{2\alpha_2}\left(1\pm\sqrt{1- 4\alpha_0\alpha_2}\right)\, . 
\end{equation}
In order to have the boundary of the asymptotic AdS$_{d}$ conformal to $\left( d-1 \right)$-Minkowski space with speed of light equal to $1$, 
$ds^2\approx \alpha_0r^2(-dt^2+d\Sigma_3^2)$, the constant $N^2$ has to be chosen as 
\begin{equation}
N^2=\frac{1}{2}\left(1\mp\sqrt{1-4\alpha_0\alpha_2}\right),\label{Nf-} 
\end{equation}
where we have the $+$ sign for the $f_-$ branch, the BB solution, while the $-$  sign has to  be used when we consider the $f_+$ branch.

In the AdS/CFT correspondence, the central charge  $c$ of the dual CFT is determined by the AdS length. 
Thus, the CFTs dual to GB gravity in both branches have central charge different from the RN case. Only in the $\a_2\to 0 $ limit the central charge 
of the $f_-$-branch coincides with that of the CFT dual to the RN theory. 
However, naive computation of the central charge in terms of the AdS length does not work in this case because of  the rescaling of the time coordinate. 
We will compute $c$ in Sect. \ref{thermodynamics}  using the scaling law of the mass and entropy as a function of the temperature.

For $\a_2<0$, only the $f_-$ branch is asymptotically AdS. 
Conversely, the $f_+$ branch describes a spacetime which is asymptotically de Sitter (dS) and can be therefore relevant as a cosmological solution.

\subsection{Singularities}
To determine the position of the singularities  of the spacetime we calculate the scalar curvature for both the $f_{\pm}$
branches:
\begin{equation}
 R^{(\pm)}=\mp \frac{1}{2}\frac{\beta r^2(20r^{10}+30\sigma r^6-31\rho r^4+6\sigma^2r^2-9\rho\sigma)\pm 20r^3(r^6+r^2\sigma-\rho)^{3/2}+2\beta\rho^2}
 {\alpha_2r^3(r^6+\sigma r^2-\rho)^{3/2}}\label{scur},
\end{equation}
where the $\pm$ sign refers respectively to the $f_{\pm}$ branches.   
To simplify expressions we used (here and after) the following notation 
\be\label{pp}
\beta= \sqrt{1-4\a_0\a_2},\quad \sigma= \frac{4 \a_2\oc M}{\beta^2},\quad \rho=  \frac{16\pi G_N\a_2 Q^2}{3\beta^2},
\quad e=\frac{1}{\beta^2}-1=\frac{4\a_0\a_2}{\beta^2},\,  Y=r^2.
\ee
There are curvature singularities at $r=0$ and 
at the zeroes of the argument of the square root in Eq. (\ref{scur})  (branch-point  singularities).
The position of the physical singularities of the spacetime is therefore determined  by the pattern of zeroes of the function $g(Y)$,  
with  
 \be\lb{g1}
  g(Y)=Y^3+\s Y-\rho.
  \ee
The singularity will be located at the biggest positive zero $Y_1$ of  $g(Y)$ or at $r=0$ when $g(Y)$ has no zeroes for positive  $Y$.  
The singularity at  $Y=Y_1$ is a branch point singularity.
The pattern of zeroes of $g(Y)$  is determined by the signs of the coefficients $\rho,\sigma$ and the discriminant 
$\Delta = \left(\frac{\rho}{2}\right)^2+\left(\frac{\sigma}{3}\right)^3$. 
\begin{itemize}
    \item For $\sigma>0$, the function $g(Y)$ is a monotonic increasing function of $Y$ with a single zero 
    which, depending on the sign of $\rho$, will be positive $Y=Y_1$  ($\rho>0$) or negative  ($\rho<0$). 
    The physical spacetime  singularity will be therefore located at $r=\sqrt{Y_1}$ for $\rho,\sigma>0$ and at $r=0$ for $\rho<0,\,\sigma>0$.
    \item For $\sigma<0$, the function $g(Y)$ is an oscillating function with a maximum at negative $Y$ and a minimum at positive $Y$, 
    it may therefore have one, two or three zeros.  For $\sigma<0,\,\rho>0$, $g(Y)$ has at least a positive zero.  
    For $\sigma<0,\,\rho<0$ we have   a positive zero for  $\Delta \le 0$ and no positive zeros for $\Delta>0$. For  
    $\Delta=0$ we have a double zero of $g(Y)$ so that $Y_1$ is not anymore a  branch point singularity. In this latter case the singularity is at $r=0$. 
\end{itemize}
Summarising, the physical singularity is always located at $r=\sqrt{Y_1}$  unless $\sigma>0,\rho<0$ or $\sigma<0,\rho<0, \Delta\ge0$ in which 
case the singularity is at $r=0$.

\subsection{$f_-$-Branch}
In this subsection, we study in detail the horizons of the $f_-$-branch, solution of Eq. \eqref{f}, describing the GB black brane.
%In this subsection we study in detail the horizons of the $f_-$-branch of the solution \eqref{f} describing the GB black brane.
In general the BB will have an inner  ($r=r_-$) and outer ($r=r_+$) event horizon. The BB becomes extremal when $r_+=r_-$. 
Using the notation \eqref{pp}, \eqref{g1}, one finds that the necessary condition for the existence of the BB  is the positivity of 
the argument in the square root in Eq. \eqref{f}, \textit{i.e.},
 $ g(Y)\ge 0$.
The position of the event horizon(s) is determined by the {\sl positive} roots of the cubic equation 
\be\lb{d1}
h(Y)=eY^3-\s Y+\rho=0.
\ee
We will first consider the case $\a_2>0$, which corresponds to $\s,\rho,e>0$ (since also  $\a_0>0$). 
The condition for the existence of real roots of the function $h(Y)$ can be easily found: 
The function $h(Y)$ has a maximum (minimum) for, respectively
\begin{equation}
Y=Y_{M,m}=\pm \sqrt{\frac{\s}{3e}}= \pm \sqrt{\frac{\oc ML^2}{3}}
\end{equation}
also, $h(Y=0)=\rho>0$, 
hence the cubic equation (\ref{d1}) always has a negative root. 
The existence of other roots is determined by the sign of $h(Y_m)$. 
We will have two (one) positive roots hence a BB with two (one)  event horizons for  $h(Y_m)\le 0$, \textit{i.e.}, for   
 \be\lb{bb}
 \rho\le \frac{2}{3}\s \sqrt{\frac{\s}{3e}}.
 \ee
Using Eq. \eqref{pp}, the previous inequality can be written in terms of the charge $Q$ and the effective mass $M$  
and gives {\sl the same}
Bogomol'nyi-Prasad-Sommerfield (BPS) bound \eqref{bbrn} found in the RN case. 
However, the BPS bound is modified when we instead express it in terms of the ADM mass: 
 \be\lb{bbrn1}
 M_{ADM}^3\ge 12 N^3 \pi^2\frac{G_N^2Q^4}{\oc^3L^2}.
\ee
 
When the bound is saturated, the inner and outer horizon merge at $r_-=r_+$, 
the BB becomes extremal, and the solution describes a soliton. 
The striking feature of the BPS bound \eqref{bb} is that the {\sl BPS bound of 5$d$ Gauss-Bonnet  
BB does not depend on the Lovelock coupling constant, and it is exactly the same 
one gets for  GR ($\a_2=0$), \textit{i.e.}, for the 5$d$ Reissner-Nordstr\"{o}m BB.}
When $M$ does not satisfy the inequality \eqref{bb}, the spacetime describes a naked singularity.  
For  $\a_2>0$, the condition $M>0$ implies $\sigma,\rho>0$ and the function $g(Y)$ is a monotonic increasing function which cuts the $Y$-axis 
at the point $Y_1$, and, in view of the previous discussion, it also gives the position of the singularity.
Since, the position of the event horizon $Y_h$ is determined by the  equation 
\begin{equation}
\beta \sqrt {g(Y_h)}=Y\sqrt{Y_h}\,, 
\end{equation}
from which follow that $g(Y_h)>0$ hence $Y_h>Y_1$,
this checks that in the region where  the bound \eqref{bb} holds the condition $g(Y)>0$  
is always satisfied and that the physical singularity is always shielded by two (one in the extremal case)  event horizons.

The behaviour of the metric function $f_-$ for $\a_2,M>0$ and  selected values of the other parameters  is  shown in Fig. \ref{fig:1}. 
The solid red, green and brown lines describe respectively a naked singularity, extremal and two-horizon BB geometry. 
The solid blue line represents a zero-charge, BB solution with single horizon.
\begin{figure}
  \includegraphics[width=\linewidth]{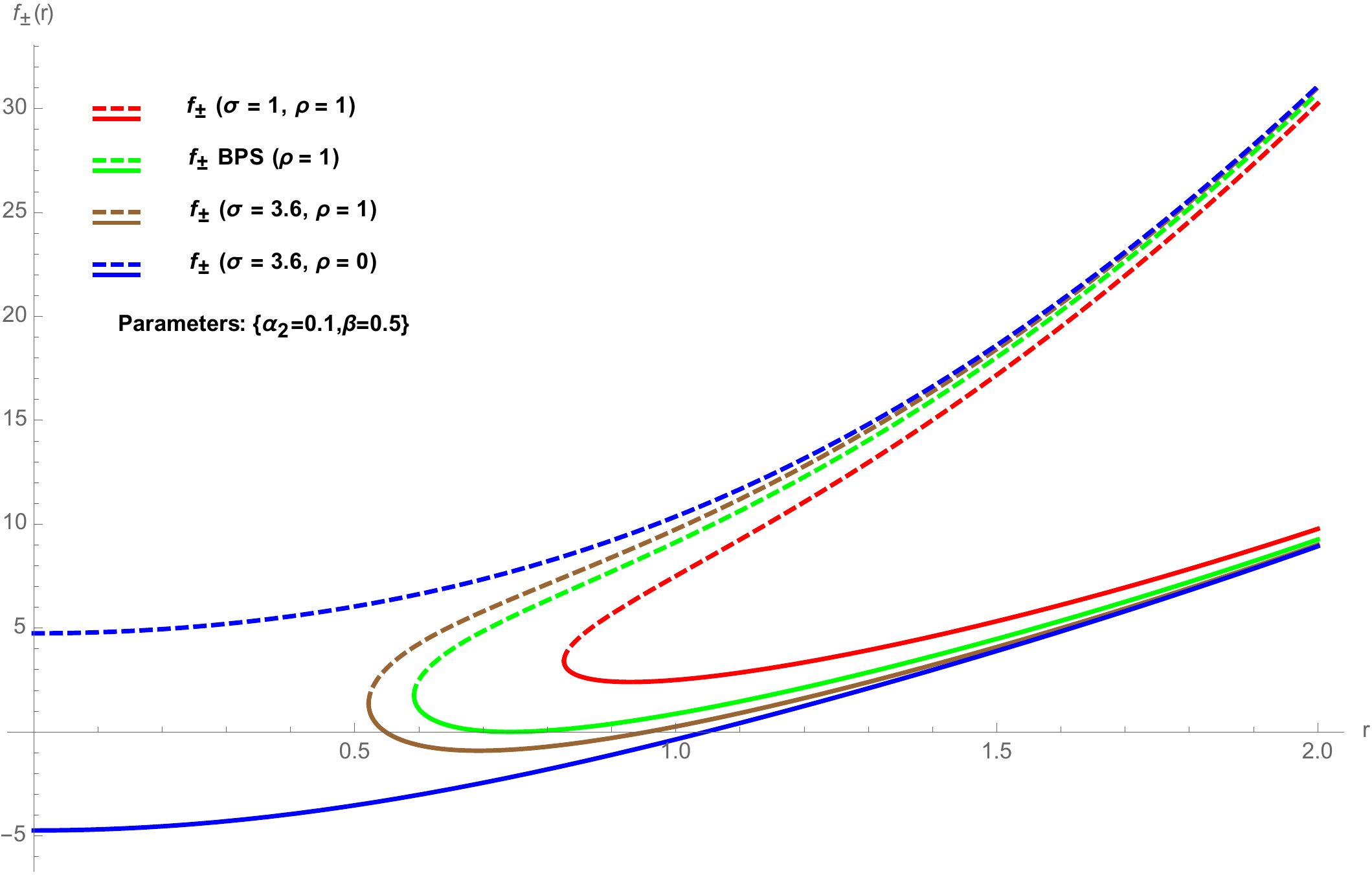}
  \caption{ Behaviour of the metric functions $f_{\pm}$ for $\a_2,M>0$ and  selected values of the other  parameters. 
  The dashed (solid)  lines describe the $f_+$ branch ($f_-$-branch). The red, green, brown and blue solid lines  describe 
  respectively a naked singularity, an extremal, two-horizon and vanishing charge BB geometry. The corresponding dashed  
  lines  describe spacetimes with a naked singularity.}
  \label{fig:1}
\end{figure}

The case $\a_2<0, M>0$ gives exactly the same BPS  bound. Now, we have $\s,\rho, e<0$. The function $h(Y)$ in Eq. (\ref{d1}) 
 always has a negative root and a minimum (maximum)  for  
\begin{equation}
Y=Y_{m,M}=\pm \sqrt{\frac{\s}{3e}}= \pm \sqrt{\frac{\oc ML^2}{3}}\,. 
\end{equation}
The conditions for the existence of two positive roots become $|\rho|\le \frac{2}{3}|\s| \sqrt{\frac{\s}{3e}}$ leading to the same BPS bound (\ref{bb}). 
However, there is a crucial difference from the $\a_2>0$ case. 
When $\a_2<0$, the condition $M>0$ implies $\sigma,\rho<0$. Taking into account that  $|e|<1$ owing to \eqref{v1}, we see that the condition $\Delta<0$ implies the 
BPS bound \eqref{bb}. This means that the two horizons are separated by a region in which the solution does not exist. 
The spacetime breaks into two disconnected parts. The physical part, having an asymptotic AdS region, describes a BB with singularity 
shielded by a {\sl single} event horizon. 
The behaviour of the metric function $f_-$ for $\a_2<0$ and  selected values of the other parameters  is  shown in Fig. \ref{fig:2}. 
The solid red, green and brown lines describe respectively a naked singularity, extremal and single-horizon BB geometry. 
The solid blue line represents a zero-charge, BB solution with  horizon.
\begin{figure}
  \includegraphics[width=\linewidth]{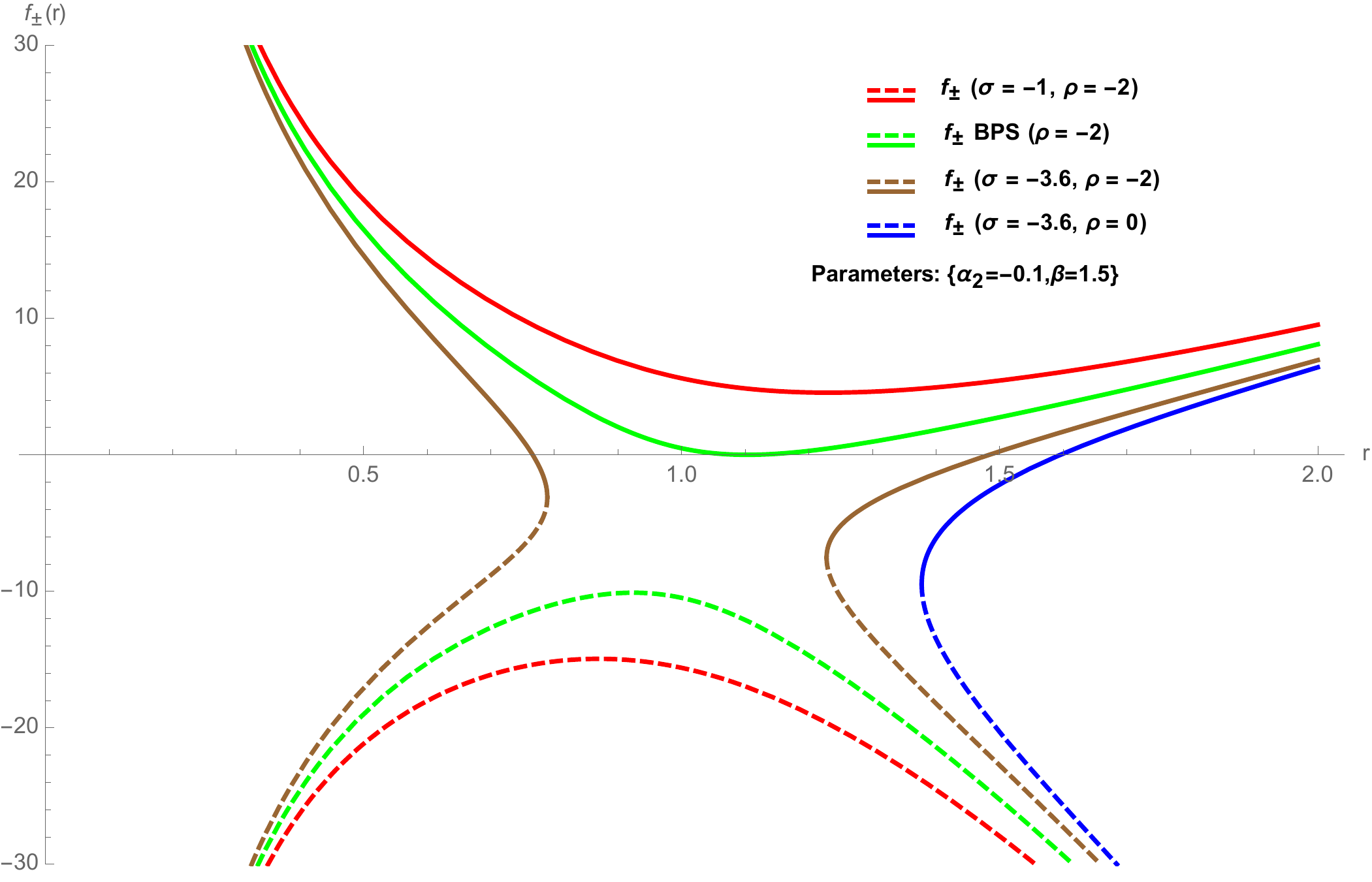}
  \caption{ Behaviour of the metric functions $f_{\pm}$ for $\a_2<0, M>0$ and  selected values of the parameters. 
  The dashed (solid)  lines describe the $f_+$ branch ($f_-$ branch). The red, green, brown, blue solid lines  
  describe respectively a naked singularity, an extremal,  single-horizon, vanishing charge BB geometry. 
  The corresponding dashed  lines  describe cosmological solutions with a singularity  which approach asymptotically to the dS spacetime.}
  \label{fig:2}
\end{figure}

\subsubsection{ Near horizon extremal solution}
\lb{sect:nhg}

When the bound \eqref{bb} is saturated, the BB becomes extremal and the metric function \eqref{f} has a double zero at  
\be\lb{t1}
Y_h=Y_m=\sqrt\frac{\s}{3e}=\sqrt{\frac{\oc ML^2}{3}},
\ee
thus, the solution $f_{-}$ can be factorized as
\be\lb{em}
f_-^{(ex)}(Y)= \frac{e\beta^2}{2\a_2}\,\frac{(Y+2Y_m)(Y-Y_m)^2}{Y^2+\beta\sqrt{Y^4+\s Y^2-\rho Y}}.
\ee 
This solution represents the extremal GB soliton.

Let us now consider the near-horizon geometry. In this regime, the solution \eqref{em} can be expanded around $r=r_0=\left(\frac{\s}{3e}\right)^{1/4}$.
At the leading order the Einstein branch reads
\be\lb{nh}
f_-^{(ex)}(r)= 12 \a_0 (r-r_0)^2.
\ee
Translating the radial coordinate $r\to r+r_0$ and rescaling the time coordinate as $t\to t/N$ we get the extremal, near-horizon geometry:

\be\lb{nhg}
ds^2= -\left(\frac{r}{l}\right)^2 dt^2+\left(\frac{l}{r}\right)^2 dr^2+ \left(\frac{r_0}{L}\right)^2 d\Sigma^2_3,\qquad l^2=\frac{1}{12\a_0}\,.
 \ee
\textit{i.e.}, AdS$_2 \times R_3$ with the AdS$_2$ length $l$ being determined uniquely by $\a_0$. Thus,
the extremal near-horizon geometry does not depend on $\a_2$ 
and fully coincides with the extremal  near-horizon 
geometry \eqref{nhg1} one gets in  the RN case.

%--------------------------------------------------------------------
\subsection{Near horizon metric as exact Solution of equations of motion}\label{nearhorizonexactsolution} 
In this section, we will show that the near-horizon solution given in Eq. \eqref{nhg}  is an exact solution of the equations of motion (EOM). 
For the GB case, Eqs. \eqref{eq:Graveq} read 
%In fact, as one can see from  Eq. \eqref{eq:Graveq} the latter are
\begin{equation}
\begin{split}
R_{ab}-\frac{1}{2}Rg_{ab}=&\frac{6}{L^2}g_{ab} +8\pi G_N \left(F_{a c} {F_b{}^{\,c}} -\frac{1}{4}g_{a b} F_{c d}F^{c d}\right)\\
&+\frac{\alpha_2}{2}g_{ab}\left(R_{cdef}R^{cdef}-4R_{cd}R^{cd}+R^2\right)\\
&+\alpha_2\left(-2RR_{ab}+4R_{ac}R_{\ b}^{c} +4R_{cd}R_{\ a\ b}^{c\ d}-2R_{acde}R_{b}^{\ cde}\right)\, .\label{EOM} 
\end{split}
\end{equation}
We note that, since the Eq. \eqref{nhg} describes a spacetime with AdS$_2\times R_3$ geometry,  
the contribution  to the curvature tensors coming from the  planar geometry $R_3$ vanishes. 
Thus, the EOM includes only the contribution of the AdS$_2$ part of the metric 
which is a two dimensional maximally symmetric space.

For a generic  $n$-dimensional maximally symmetric space with $R=\Lambda$ the  two terms in Eqs. \eqref{EOM}, that are quadratic in the curvature 
tensors, are given respectively by
\begin{equation}\lb{fds}
 \a_2\Lambda^2 \frac{(n-2)(n-3)}{2n(n-1)},\quad \quad  -2 \a_2\Lambda^2 \frac{(n-2)(n-3)}{n^2(n-1)}.
 \end{equation}
These relations are consequence of the fact that the GB term in the action is topological for $d=4$ and identically vanishes for 
$d=2$ and $d=3$. 
The previous equations imply that in the case of the AdS$_2\times R_3$ geometry,  
the contributions given by the GB terms to the EOM vanish; therefore, the near horizon metric \eqref{nhg} is an \textit{exact} solution of both Einstein and GB EOM. 
In particular, the latter reduces to the usual  Einstein-Maxwell equations in 5$d$.\\

Summarising, we have seen that the AdS$_2 \times R_3$ geometry is not only a near horizon approximation but it is an exact solution of the 
field equations of GB-Maxwell gravity. 
The presence of two exact extremal  solutions (the extremal soliton interpolating through a throat region the AdS$_2 \times R_3$ geometry with 
the asymptotic AdS geometry and the AdS$_2 \times R_3$ geometry itself) is a typical feature of extreme black branes describing  BPS states 
(see e.g. Refs. \cite{Maldacena:1998uz, Cadoni:2000hg}). 
The two solutions correspond to two different  extremal limits. As we will see in Sect. \ref{thermodynamics}, the presence of two different extremal, 
exact, solutions give rise
to a non-trivial extremal thermodynamic behaviour.

%----------------------------------------------------------------------------------------------------

\subsection{$f_+$ Branch}
This branch does not describe a BB but a
spacetime with a  singularity for every value of the parameters $Q\neq 0, M\neq  0$.  
Depending on the value of the parameter $\a_2$ we have either a spacetime with a naked 
singularity (for $\a_2>0$) or a cosmological, asymptotically de Sitter (dS) solution with a singularity (for $\a_2<0$.) 
This follows from the  above discussion of the singularities of the scalar curvature  \eqref{scur}.
In the $f_+$ branch  the spacetime always has a singularity, which can be located  at $r=0$ or $r=\sqrt{Y_1}$ 
depending on the values of the parameters.
This is consistent with the results of  Ref. \cite{Boulware:1985wk}, according to which   the $f_+$ branch is unstable and contains ghosts\footnote{
In principle, one could have hoped to have a regular spacetime when the function $g(Y)$ has a double zero at positive $Y$. 
In fact in this case the branch point singularity is removed and if the spacetime  in the region $Y_1\le Y<\infty$ 
is geodesically complete we have regular, 
solitonic geometry. The function $g(Y)$ has a double zero at positive $Y$ for $\sigma,\rho<0,\,\Delta=0$, 
but unfortunately the spacetime cut at $Y=Y_1$ thus it is not geodesically 
complete. }. 

For $M,\a_2>0$, the metric functions for the $f_+$ branch are the dashed lines shown  in Figs. \ref{fig:1}. 
An interesting, peculiar feature is that in this case, all the solutions of the $f_-$ branch are continuously connected with the solution  
of the $f_+$-branch passing trough the singularity. This feature has a simple analytic explanation. 
In the cases under consideration the singularities are the zeros of the function $g(Y)$ and when $g(Y)=0$ then $f_+=f_-$. 
This fact can have interesting holographic implications: we have two CFTs with different central charges connected through  
the same singularity.

For $M>0$ and $\a_2< 0$,  the $f_+$ branch  describes a cosmological solution with a singularity. The corresponding metric functions  are shown 
(dashed lines) 
in Fig. \ref{fig:2}. Also in this case  the solutions of the $f_-$ branch with an horizon  are continuously  
connected with the solution  of the $f_+$-branch passing trough the singularity. We have now an asymptotically AdS solution continuously connected
through a cosmological singularity to a late de Sitter geometry. On the other hand, the solutions of the $f_-$ branch describing a naked singularity are 
disconnected from the cosmological solutions.

For $\a_2,M<0$,  the $f_+$ branch describes a cosmological solution  with a singularity with late de Sitter behaviour, 
whereas the $f_-$ branch describes an 
asymptotically AdS spacetime with a naked singularity.  
However, here the two branches are disconnected. The metric functions for this case  are shown
in Fig. \ref{fig:3}. 

It should be stressed that in the $Q=0$ case, the $f_+$ branch has ghosts in the spectrum \cite{Boulware:1985wk}.
We naturally expect this to extend to the charged case and is consistent with the intrinsic instability of these branch of  solutions connected with 
the presence of naked singularities.

 \begin{figure}
  \includegraphics[width=\linewidth]{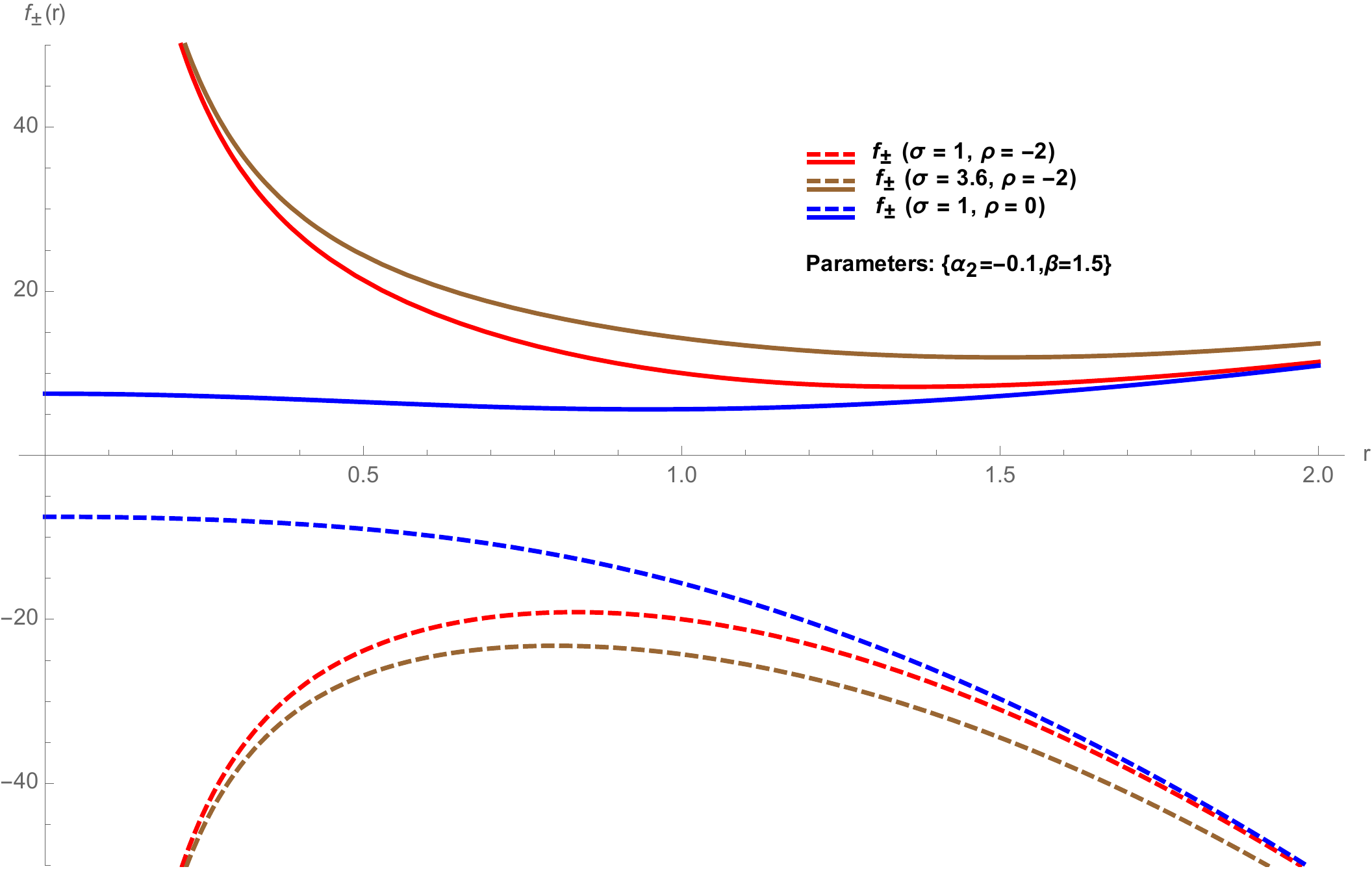}
  \caption{Behaviour of the metric functions $f_{\pm}$ for $\a_2,M<0$ and  selected values of the other parameters. 
  The dashed (solid)  lines describe the $f_+$ branch ($f_-$branch). The solid lines describe spacetimes with naked singularities, 
  whereas the dashed lines    describe cosmological, asymptotically dS solutions with a   singularity.}
  \label{fig:3}
\end{figure}
 
%----------------------------------------------------------
\section{Charged GB black brane thermodynamics}\label{thermodynamics}
In this section, we will study the thermodynamics of the GB BB solutions, \textit{i.e.,} solutions in the $f_-$ branch 
and make a comparison with the Reissner-Nordstr\"{o}m black branes.\\
The effective  thermodynamic potentials $M=M_{ADM}/N,S,\Phi$ and the temperature $T=T_H/N$ can be written 
as functions of the horizon radius $r_+$ and the charge $Q$  by specializing  
Eqs. \eqref{eq:Mass},\eqref{T},\eqref{eq:Entropy} to  $d=5$. We obtain the following equations
\bea\lb{MTSF}
 M&=&\frac{r_+^4}{\omega_5L^2}\left(1+\frac{4\pi}{3}\frac{G_NQ^2L^2}{r_+^6}\right),\quad 
 T=\frac{1}{\pi L^2}\left(r_+-\frac{2\pi G_NQ^2L^2}{3r_+^5}\right),\nonumber\\
  S&=&\frac{V_3}{4G_N}\left(\frac{r_+}{L}\right)^3,\quad 
  \Phi=\frac{V_{3}}{2 L^{3}}\frac{Q}{r_+^{2}}\, ,
\eea
 that satisfy the first principle  $dM=TdS+\Phi dQ$.
As pointed out in Sect. \ref{sect:ub}, because of the universality of the  thermodynamic behaviour, the  thermodynamic 
relations \eqref{MTSF} hold for both for the charged GB and the RN BB. 
The only difference is that for the GB brane,  with metric function \eqref{f}, $M$ and $T$ 
are the effective  parameters whereas in the RN case $M=M_{ADM}$ and $T=T_H$.

In order to have a clear and complete  description  of the GB BB thermodynamics, one should eliminate $r_+$ from the Eqs. \eqref{MTSF} 
and write $M(T,Q), S(T,Q)$. This parametrization cannot be done in analytic form because we have to solve a  $6^{th}$ grade equation in $r_+$. 
%In the next section 
Thus, we will derive the explicit scaling behaviour of $M$ and $S$ as a function of the temperature in the large and small $T$ limit.
These relations will shed light on  the holographic interpretation of  the solutions. 
%
%\subsection{Scaling behaviour}
%
The functions $M(T,Q)$ and $S(T,Q)$ can be obtained in implicit form  by using the second equation in \eqref{MTSF}
as an implicit definition of the function $r_+(T,Q)$, and they read %Moreover, the same equation can be used  to eliminate $Q$.  We get 
\begin{equation}
 M(T,Q)=\frac{r_+^3}{\omega_5L^2}\left(3r_+-2\pi L^2T\right),\quad\quad S(T,Q)=\frac{V_3}{4G_N}\left(\frac{r_+}{L}\right)^3\, .
\label{MassT}
\end{equation}
%
%where the $r_+=r_+(T,Q)$ is  defined implicitly by the second Equation in  Eq.(\ref{MTSF}).
%
Let us now consider separately the two limits of interest: $T\to\infty$ and $T\to 0$.

\subsection {Large temperature}
\lb{sect:LT}
The limit $T\to\infty$ corresponds to large radius BB, \textit{i.e.}, $r_+\to \infty$. In this regime, the temperature scales linearly with 
$r_+$
\begin{equation}\lb{kli}
 T\simeq\frac{r_+}{\pi L^2} 
\end{equation}
and, at the leading order, we get for $M$ and $S$ 

\begin{equation}\lb{MSAS}
 M=\frac{3V_3L^3}{16\pi G_N}\left(\pi T\right)^4, \quad \quad S=\frac{V_3 L^3}{4G_N}\left({\pi T}\right)^3.
\end{equation}
This is exactly the scaling behaviour one expects for a UV  fixed point described by a CFT$_{4}$.

Because of the universality of the thermodynamic behaviour, the relations \eqref{MSAS} hold for both the RN and the GB BB. 
In the former case, Eqs. \eqref{MSAS} hold when $M=M_{ADM}, T=T_H$, in the latter when $M,T$ are given by 
the effective values in Eq. \eqref{eff}. Thus, for the GB BB, mass and entropy acquire a $1/N^3$ factor.  

The central charge $c$ of the associated CFT is determined  by the proportionality factor and  can be easily calculated. 
In the case of the RN BB, when $M=M_{ADM}$ and $T=T_H$ in Eq. \eqref{MTSF},  we have $c\propto L^3/G_N$. 
On the other hand, in the GB BB case, we have seen that the same thermodynamic relations  \eqref{MTSF} hold for 
 $M,T$ given by the effective values in Eq. \eqref{eff} %effective mass and temperature 
and we will get from Eqs. \eqref{MSAS} 
\be\lb{central}
c\propto \frac{L^3}{N^3G_N}.
\ee

\subsection {Small temperature}
The  $T\to 0$  thermodynamic behaviour corresponds to extremal BBs in which the BPS bound \eqref{bbrn1} is saturated.
This is achieved at  non vanishing, constant value of the BB radius
\begin{equation}\label{r0}
 r_+=\left(\frac{2\pi G_NL^2Q^2}{3}\right)^{1/6}\equiv r_0
\end{equation}
that corresponds, as expected for BPS states, to the extremal brane  $T=0$ state with  non vanishing mass and  entropy given by
\be\lb{MSBPS}
M_{ext}=\frac{3r_0^4}{2 \omega_5L^2},\quad\quad  S_{ext}=\frac{V_3}{4G_N}\left(\frac{r_0}{L}\right)^3.
\ee
We can now expand in Taylor series the temperature near $r_0$ to obtain 

\begin{equation}\label{smallT}
 T\simeq \frac{3}{\pi L^2}\left[2(r_+-r_0)-\frac{5}{r_0}(r_+-r_0)^2\right], 
\end{equation}
and  the behaviour of $M$ and $S$ near $T=0$ is of the form
\begin{equation}\lb{MSNE}
 M- M_{ext}=\frac{2r_0^2}{3\omega_5}(\pi LT)^2 + {\cal O}(T^4),\quad 
 S-S_{ext}=\frac{\pi r_0^2 V_3}{8G_NL} T + {\cal O}(T^2)\,.
 \end{equation}

Again,  universality of the thermodynamic behaviour imply that the relations in Eq. \eqref{MSNE} hold both for the RN  and for the GB
BB. For the RN case, the relations take  the same form  with $M=M_{ADM}$ and $T=T_H$. 
For the GB case, when we express the relations \eqref{MSNE} 
in terms of ADM mass and Hawking temperature we get  
\begin{equation}
\begin{split}
M_{ADM}&= N M_{ext}+\frac{2r_0^2}{3 N \omega_5}(\pi LT_H)^2 + {\cal O}(T^4)\,\\
 S&= S_{ext}+\frac{\pi r_0^2 V_3}{8N G_NL} T_H + {\cal O}(T^2).
\end{split}
\end{equation}

%{\mar
\subsection{Excitations near extremality and near-horizon limit}
\lb{sect:ex}
%%%
%%
An important feature of the RN BB, which in view of the previous results extends to the charged GB BB, is  
that the semiclassical analysis of its thermodynamic behavior breaks down near extremality \cite{Maldacena:1998uz,Cadoni:2000hg}. 
In fact, the energy of an Hawking radiation mode is of order $T_H$ and the semiclassical description breaks down when this energy 
is comparable with the energy above extremality $M- M_{ext}$ given by  Eq. \eqref{MSNE}.
This results in an energy gap for excitations above extremality \cite{Maldacena:1998uz}, which in the case under consideration is  
$E_{gap}\sim (N\omega_5)/L^2r_0^2$. The fact that the extremal limit is singular, can be also understood in geometrical terms. 
It has been observed that  at extremality the geometry  splits into two spacetimes:  
an extremal black hole and a disconnected AdS space \cite{Carroll:2009maa}.

The presence of this energy gap has important consequences for what concerns the spectrum of BB excitations near extremality.
In particular,  whereas in the extremal case  the near-horizon geometry is given, as shown in Sect. \ref{sect:nhg}, 
by AdS$_2\times R_3$, finite energy excitations of AdS$_2\times R_3$ are suppressed. 
Analogously  to the RN case in 4$d$  \cite{Maldacena:1998uz}, one can consider   
near-horizon limits not restricted to  zero temperature and excitation energy. 
These limits are obtained by letting the 5$d$ Planck length $L_P$ go to zero, holding fixed some of the other 
physical parameters of the  BB (energy, charge and temperature).

\section{Shear viscosity to entropy ratio}\label{General_eta/s}
The \textit{universality} of the shear viscosity $\eta$  to entropy density $s$ ratio  
for Einstein-Hilbert  gravity represents a very important result of the  gauge/gravity correspondence.  
First found for  the hydrodynamic  regime of  the QFT dual to   black branes and black holes of the Einstein-Hilbert 
theory \cite{Policastro:2001, KSS:2003, KSS:2005}, the   KSS bound $\eta/s\geq1/4\pi$ has been extended to a variety  
of cases. These include Einstein-Hilbert gravity with all possible matter terms in the action, hence, among others the QFT dual to Reissner-Nordstr\"{o}m 5$d$ gravity \cite{KSS:2003, KSS:2005} and  the important case of the quark-gluon plasma
(see e.g. \cite{Cremonini:2011}).  
It has been also conjectured that the KSS bound  holds for any fluid in nature. For a detailed  discussion on the shear viscosity to entropy 
ratio see Refs. \cite{Policastro:2001, KSS:2003, KSS:2005, Ge-Sin:2009, Ge:2008ni,
Brigante-Myers:2008,Cai:2008ph, Cremonini:2011,Cai:2009zv,Astefanesei:2010dk}.

The KSS bound seems to lose its universality when one introduces,  in the Einstein-Hilbert action, higher powers of the curvature  tensors. 
This is, for instance, the case of Lovelock (and Gauss-Bonnet) gravity we are discussing in this paper. 
In particular, the KSS bound  depends on the  coupling constant for the higher curvature terms \cite{Brigante-Myers:2008}. 

Following the notation of \cite{Brigante-Myers:2008}, we rewrite  the GB BB solution  \eqref{f} as follows
\begin{equation}\lb{me}
 f_-=\frac{r^2}{2\lambda L^2}\left[1-\sqrt{1-4\lambda\left(1-\frac{\omega_5ML^2}{r^4}+\frac{4\pi}{3}\frac{G_NQ^2L^2}{r^6}\right)}\right],
\end{equation}
where $\alpha_0\alpha_2=\alpha_2/L^2=\lambda$.
In 5$d$ Gauss-Bonnet gravity, the shear viscosity to entropy ratio  is \cite{Brigante-Myers:2008}
\begin{equation}\label{GBKSS}
 \frac{\eta}{s}=\frac{1}{4\pi}\left(1-4\lambda\right).
\end{equation}
 The KSS bound still holds if $\lambda\le 0$ but is violated for  $0<\lambda\le 1/4$  
(the upper bound follows from  Eq. \eqref{v1}).
The dependence of the bound from the coupling constant $\lambda$ makes the bound not anymore universal as in the Einstein-Hilbert theory. 
In terms of the dual gauge theory,  
the curvature corrections  to the  Einstein-Hilbert action correspond to finite ${\cal{N}}$ and $\lambda_{tH}$ effects. 
It has been argued that the universality of the KSS bound strictly holds in the limit ${\cal{N}}\to\infty$ whereas, in general, finite ${\cal{N}}$ effects will give lower bounds for $\eta/s$ \cite{Kats:2007mq}.  

A crucial issue  is that the relation (\ref{GBKSS}) seems to allow for arbitrary violations  of the KSS bound. However, consistency of the QFT  dual to  bulk GB gravity as a relativistic field theory constrains  the allowed values of $\lambda$. 
For instance, in 
\cite{Brigante:2008gz,Buchel:2009tt,Hofman:2009ug} it was found that causality and  positivity of the energy for   
the dual QFT describing the  Gauss-Bonnet plasma require   $-7/36<\lambda<9/100$ implying   $4\pi\eta/s> 16/25$ , a bound lower then the KSS bound.
On the other hand, the hydrodynamic description of the dual GB plasma is valid in the IR regime, \textit{i.e.}, for $\omega,k<<T$, whereas causality 
is determined by the propagation of modes in the $\omega,k>T$, UV regime.  Thus, the existence of lower bounds for  $\eta/s$
implies a higher non-trivial relationship between the transport properties in the IR and causality requirements in the UV regime of the QFT dual to GB gravity.

%Moreover, r
Recent investigations have shown that if translation symmetry is broken in the IR then one may have strong violation of 
the KSS bound even in the context of Einstein gravity, in the form of $\eta/s\sim T^{2\nu},\,\nu\le1$ \cite{Jain:2014vka,Jain:2015txa,Hartnoll:2016tri}. 
Although, for these backgrounds, the breaking of translational invariance prevents an hydrodynamical interpretation of the viscosity,
this  behaviour of $\eta/s$ is clearly related to the emergence of extremely interesting physics  in the far IR. 

A way to shed light on these questions is to investigate the behaviour of $\eta/s$ in the case of a gravitational bulk background 
for which there is a temperature  flow of $\eta/s$. 
The charged GB BB represents a nice example  of this behaviour, particularly in view of the universality of the IR  AdS$_2\times R_3$ fixed point. This will be the subject of the next three subsections.

\subsection{$\eta/s$ for the charged GB black brane}
\lb{sect:ppp}

A standard  way to calculate the shear viscosity for a QFT  is by using the Kubo formula  for the transverse  momentum conductivity
\be\lb{kf}
\eta=\lim_{\omega\to 0} \frac{1}{\omega} Im G^R_{xy},
\ee 
where $G^R_{xy}$ is the retarded Green function for the $T_{xy}$ component of the stress-energy tensor.

The application of the usual AdS/CFT procedure for the computation of correlators gives for the $U(1)$-charged Gauss-Bonnet 
black brane in five dimensions   \cite{Ge:2008ni,Cai:2008ph}
\begin{equation}\lb{visc}
 \eta=\frac{s}{4\pi}\left[1-4\lambda\left(1-\frac{a}{2}\right)\right]
\end{equation}
where $a=\frac{4\pi}{3}\frac{G_NQ^2L^2}{r_+^6}$,  and $s$ is  
 the entropy density $S/V$ following from (\ref{MTSF}).

A drawback of the usual computation of the  shear viscosity  is that it does not work in the extremal $T=0$ case 
because the metric function has a double zero at the horizon. For this reason, $\eta$ in the case of extremal BB cannot be simply computed by taking the $T_H= 0$ limit in Eq. (\ref{visc}). Building on  \cite{Faulkner:2009wj},  a method of dealing with this problem  has been developed in \cite{Edalati:2009bi}. 
Recently, a very simple and elegant formula for computing correlators of the form (\ref{kf}) in QFTs dual to a gravitational bulk theory 
has been proposed in  \cite{Lucas:2015vna} (see also
 \cite{Hartnoll:2016tri,Davison:2015taa}). This method also works in the extremal case; thus, in the following, we will use it to 
 compute $\eta$ for the charged GB BB.

Considering perturbations $g_{ab}=g^{(0)}_{ab}+h_{ab}$ of the background (\ref{me}), at the linear level 
the field equations (\ref{EOM}) give for the $h^y_x(t,r)=\phi(r)e^{-i\omega t}$ component of the perturbation
\be\lb{pert}
\partial_r\left[\sqrt{\gamma(r)}f_-(r)F(r)\partial_r \phi\right]+ \omega^2\frac{\sqrt{\gamma(r)}F(r)}{N^2f_-(r)}\phi=0,
\ee
where $\gamma(r)= (r/L)^3$ is the determinant of the spatial metric, $f_-(r)$ is given 
by Eq. \eqref{me} and $F= N\left(1- \frac{\lambda L^2}{r} \partial_rf_-(r)\right)$. 
Notice that in the background \eqref{me}, the component   $h^y_x$ decouples from the other perturbation modes.

Let us first consider the non extremal black brane. The extremal case will be discussed in Sect. \ref{sect:ex1}.  
Following Ref. \cite{Lucas:2015vna} we now denote with $\phi_0(r)$ the time independent solution of \eqref{pert} which is 
regular on the horizon $r=r_+$ and such that $\phi_0\to1$ as $r\to\infty$. The other linearly independent solution $\phi_1(r)$ of Eq. \eqref{pert} 
behaves as $1/r^4$ for  $r=\infty$ and can be computed using the Wronskian method,
\be\lb{ss}
\phi_1= \phi_0\int_r^\infty \frac{dr}{\phi_0^2\sqrt\gamma Ff_-}.
\ee
Expanding near the horizon $r=r_+$ we get at leading order 
\be\lb{exp}
\phi(r)=- \frac{1}{\phi_0(r_+)}\frac{\ln(r-r_+))}{4\pi T_H \sqrt{\gamma(r_+)}\left[1-4\lambda(1-\frac{a}{2})\right]},
\ee
where $T_H$ is the Hawking temperature of the BB and $a$ is defined as  in Eq. (\ref{visc}).
Solving now Eq. \eqref{pert} near the horizon with infalling boundary conditions and for small $\omega$, one gets at leading order in $\omega$
\be\lb{jj}
\phi(r)=\phi_0(r_+)\left(1- \frac{i\omega}{4\pi T_H}\ln(r-r_+)\right).
\ee
Comparing Eq. (\ref{exp}) with Eq. (\ref{jj}) and  expanding near the $r\to\infty$ boundary of AdS, one gets
\be\lb{ppp}
\phi(r)=1+ i \omega \phi_0^2(r_+) \sqrt{\gamma(r_+)}\left[1-4\lambda\left(1-\frac{a}{2}\right)\right]\frac{1}{r^4}.
\ee
 
The usual AdS/CFT rules for computing boundary correlators tell us that the retarded Green function is $1/(16\pi G_N)$ 
the ratio between normalizable and non-normalizable modes, so that we have
\be\lb{lll}
\eta= \frac{s}{4\pi}\phi_0(r_+)^2\left[1-4\lambda\left(1-\frac{a}{2}\right)\right].
\ee
Because $\phi_0(r)$ goes to $1$ as $r\to\infty$ and must be regular on the horizon, we  have $\phi_0(r_+)=1$ and Eq. 
\eqref{lll} reproduces correctly  the  previous result  \eqref{visc}.

Now, the second Eq. \eqref{MTSF} can be used to define, implicitly, the horizon radius as a function of the BB Hawking 
temperature and the electric charge, 
thus allowing us to write also the shear viscosity \eqref{visc} as a function of $T_H$ and $Q$
\begin{equation}\label{etaTQ}
 \eta(T_H,Q)=\frac{1}{16\pi G_N}\left(\frac{r_+(T_H,Q)}{L}\right)^3\left[1-4\lambda\frac{\pi L^2T_H}{Nr_+(T_H,Q)}\right].
\end{equation} 
In the same way, the entropy density in Eq. \eqref{MTSF} can be written as a mere function of $T_H$ and $Q$, so that we can write 
the shear viscosity to entropy ratio in the form 
\begin{equation}\label{etasTQ}
 \frac{\eta}{s}=\frac{1}{4\pi}\left[1-4\lambda\frac{\pi L^2}{Nr_+(T_H,Q)}T_H\right].
\end{equation}

It is also of  interest to write explicitly the dependence of $\eta/s$ from the  normalization constant $N$:
\begin{equation}\label{U1GBKSSN}
 \frac{\eta}{s}=\frac{1}{4\pi}\left[1-4N \pi L^2 (1-N^2)\frac{T_H}{r_+}\right]
\end{equation}

When the electric charge is set to zero, the ratio $T_H/r_+$ in Eq. \eqref{etasTQ} 
is $N/(\pi L^2)$ and $\eta/s$ reaches the value in Eq. \eqref{GBKSS}, as one expects. 
On the other hand, the dependence of $\eta/s$ on $T_H$ and $N$ in the generic case is rather puzzling.

In view of the universality of the thermodynamic behaviour of GB BB described in the previous sections one would  naively 
expect also the  shear viscosity to entropy ratio to be universal, \textit{i.e.}, that 
Eq.  \eqref{U1GBKSSN} becomes the same as in the RN case just by using the effective temperature $T=T_H/N$ instead of $T_H$.
This is not the case.   Only for $N=1$, which corresponds to $\alpha_2=0$, \textit{i.e.}, exactly the RN case,  $\eta/s$   assumes the 
universal value $1/4\pi$, while for $N$ generic  we have a quite complicated dependence on $N$ and $T_H$.
This indicates strongly that the transport features of the dual QFT in the hydrodynamic regime contain more information about the underlying microscopic theory than that contained in  the  thermodynamic description.
%
%A way to shed light on this issue is to investigate   the behaviour of $\eta/s$ at large and small $T_H$. In fact, we have seen in the previous sections that in these limits  the BB allows for  a simple thermodynamic description. We therefore expect this to be true also for the shear viscosity to  entropy  ratio. This will be the subject of the next sections.
%
An investigation on the behaviour of $\eta/s$ at large and small $T_H$ can shed light on this issue.   In fact, as we have seen in the previous sections, in these limits  the BB allows for  a simple thermodynamic description.
 We, therefore, expect this to be true also for the shear viscosity to entropy  ratio. This will be the subject of the next sections.

\subsection{$\eta/s$ in the large and small $T_H$ regime}
The behavior of  the shear viscosity \eqref{etaTQ} for large and small temperatures can be investigated 
in a way similar to that used for the BB thermodynamics.

\subsubsection{Large $T_H$}
For large $T_H$,  the Hawking temperature is given by Eq. \eqref{kli}, 
thus leading to the following expression for the shear viscosity in Eq. \eqref{etaTQ},
\begin{equation}
 \eta=\frac{1}{16\pi G_N}\left(\frac{\pi LT_H}{N}\right)^3\left(1-4\lambda\right).
\end{equation}
The shear viscosity at large $T_H$ scales as $T_H^3$. In this limit, the entropy density also depends on the temperature as $T_H^3$ (see Eq.
\eqref{MSAS}),
the shear viscosity to entropy density  ratio  approaches Eq. \eqref{GBKSS}
and reduces to the universal value $1/4\pi$ when $\lambda\to0$.
This is rather expected, because  at large $T_H$ the  contribution  of the electric charge can 
be neglected.\\

\subsubsection{Small $T_H$}

To investigate the small  $T_H$ behaviour we  invert Eq. \eqref{smallT} and we  write the horizon radius as 
\begin{equation}\lb{jjj}
r_+-r_0\simeq\frac{\pi L^2}{6N}T_H \, , 
\end{equation}
where $r_0$ is defined by Eq. \eqref{r0}. At small temperature the subleading term in the shear viscosity  scales linearly in $T_H$
\begin{equation}
 \eta\simeq\frac{1}{16\pi G_N}\left(\frac{r_0}{L}\right)^3\left[1+\left(\frac{1}{2}-4\lambda\right)\frac{\pi L^2T_H}{Nr_0}\right]\, .
\end{equation} 
 The behavior of the entropy density in the small $T_H$ regime is given by  the second equation in \eqref{MSNE}. 
Hence, in this limit, also the subleading term of the shear viscosity to entropy density  ratio scales linearly
\begin{equation}
 \frac{\eta}{s}\simeq\frac{1}{4\pi}\left[1-4\lambda\frac{\pi L^2T_H}{Nr_0}\right].
\end{equation}
The result $\eta/s=1/4\pi$ for $T_H=0$ has been already  found and discussed in the literature in the case of the RN solution
\cite{Edalati:2009bi,Faulkner:2009wj}.
It has been argued that at small temperatures,  the dual QFT behaves as  a "strange RN metal".
The optical conductivity exhibits the generic perfect-metal behaviour, but  although
we have a non-vanishing ground-state entropy, for the strange metal 
 hydrodynamics continues to apply and energy and momentum can
diffuse.

In the limit $T_H=0$, the ratio becomes $\eta/s=1/4\pi$  attaining the universal value one expects from the KSS bound.
This result is what one naturally expects in view of the fact that  at  $T_H=0$ the  near-horizon solution  of the GB brane 
gives  exactly the same  AdS$_2\times R_3$ geometry of the RN solution. 
 However, extra care is needed when one takes  the $T_H\to 0$  limit  in Eq. (\ref{etasTQ}).
Taking $T_H\to 0$ directly in Eq. (\ref{etasTQ}) is not save for several  reasons.
First, as discussed in Sect. (\ref{sect:ex}) the semiclassical description for the BB breaks down at small 
temperature when the energy gap above extremality prevents excitations with finite energy. 
Second, as noted by Cai \cite{Cai:2008ph}, although the $T_H\to 0$ limit is well defined, 
the usual computation of the shear viscosity to entropy ratio fails in the extremal case because the metric function as a double zero at the horizon.
Third,  also the computations of Sect. \ref{sect:ppp} do not hold for $T_H=0$ because  the expressions (\ref{exp}) and
(\ref{jj})  are ill defined for $T_H=0$. 
However, the general method based on \cite{Lucas:2015vna} and used in Sect. \ref{sect:ppp} for calculating $\eta$,
works also for extremal BB.

\subsection{$\eta/s$ in the extremal case}
\lb{sect:ex1}
Let us now extend  the calculations of $\eta$ described in Sect. \ref{sect:ppp} to the case of the  extremal brane.
In the extremal case   the function $f_-$ given by Eq. (\ref{me}) and its first derivative vanish when evaluated on  the horizon. We have therefore
at leading order near the horizon
\be\lb{fgr}
f_-(r_+)=f'_-(r_+)=0,\quad F(r_+)=N, \quad f_-(r)\simeq k(r-r_+)^2,
\ee
where $k$ is some  non zero constant.
Using the previous expression in (\ref{ss}) one gets
\be\lb{lkg}
\phi_1(r)= \frac{1}{k N\phi_0(r_+) \sqrt{\gamma(r_+)}} \frac{1}{(r-r_+)}.
\ee
On the other hand the near-horizon, small $\omega$ expansion  gives now
\be\lb{gjjl}
\phi(r)=\phi_0(r_+)\left[ 1+ \frac{i\omega}{kN(r-r_+)}\right].
\ee
Comparing Eqs. (\ref{lkg}) and (\ref{gjjl}), near the $r\to\infty$ boundary of AdS$_5$ we find the expansion
\be\lb{ppp1}
\phi(r)=1+ i \omega \phi_0^2(r_+) \sqrt{\gamma(r_+)}\left(\frac{1}{r^4}\right),
\ee
from which follows the shear viscosity
\be\lb{lll1}
\eta= \frac{s}{4\pi}\phi_0(r_+)^2.
\ee
Using the same argument used in Sect. (\ref{sect:ppp}) to infer that  $\phi_0(r_+)=1$,  we get  for the shear viscosity 
to entropy ratio of the extremal GB black brane the universal value 
\be
\frac{\eta}{s}=\frac{1}{4\pi} \, .
\ee
It is interesting to notice  that the universality   of $\eta/s$ for the extremal  GB BB
is a direct consequence of  the universality of the AdS$_2\times R_3$,  extremal, near-horizon geometry.
In  fact  the extremal,  near-horizon 
metric background (\ref{nhg}) does not depend  on $\lambda$. The other source for a $\lambda$- or $Q$-dependence  of $\eta$ 
is the function $F$ in Eq. (\ref{pert}). However, this contribution, hence the dependence of $\eta$ from  $\lambda$ and $Q$, 
is removed by the condition $f'(r_+)=0$, which  implies that near the horizon the two-dimensional sections of the metric behave as AdS$_2$.

To conclude, let us now discuss the global behaviour of $\eta/s$ as a function of  the temperature in order to gain some  
insight about the $\eta/s$ bounds.
Taking into account that  $r_+(T_H)$ is a monotonically increasing function, one easily finds that also  
the function $P(T_H)=\pi L^2T_H/(Nr_+)=1- 2\pi G_NQ^2L^2/(3r_+^6)$ in Eq. (\ref{etasTQ}}) 
 is a monotonically increasing function of $T_H$, with $P(0)=0$ and $P(\infty)=1$.
The global behaviour  of $\eta/s$ in   Eq. (\ref{etasTQ}) 
  therefore is ruled by  the sign of $\lambda$. For $\lambda<0$,  $\eta/s$ is a monotonically {\emph {increasing}} function of $T_H$, which raises from its
  minimum value $1/4\pi$ at $T_H=0$ to its maximum value $(1+4|\lambda|)/4\pi$ for $T_H=\infty$, in full agreement with the KSS bound.
  On the other hand,  for $0<\lambda<1/4$, $\eta/s$ is a monotonically {\emph{decreasing}} function of $T_H$, which drops  from its
  maximum  value $1/4\pi$ at $T_H=0$ to its minimum  value $(1-4\lambda)/4\pi$ for $T_H=\infty$, violating the KSS bound.

\section{Summary and outlook}\label{sec:conclusions}

In this paper, we have  discussed in detail geometrical, thermodynamic and holographic properties of charged 
5$d$ GB black branes. Although our discussion has been mainly confined to the 
GB case, we expect that most of our results can be generalised to Lovelock gravity theories in any spacetime dimensions. 

We have shown that the particular combination of GB higher curvature terms added to  the Einstein gravity action have three main effects:

(1) They introduce a new branch of brane solutions, which are however not black branes  but describe naked singularities.
The global structure of the RN geometry of Einstein gravity is preserved only for $\a_2>0$. 
For $\a_2<0$ the spacetime splits into two disconnected regions (an inner and outer region),
with the external region having a single event horizon also in the non-extremal case. 
An interesting feature is that the solutions of the two branches may be, in some cases, continuously connected one with the other through the singularity.
When  this is the case, they describe transitions of the kind: AdS$_5\to$ singularity $\to$ AdS$_5$,  
AdS$_5$-black brane $\to$ singularity $\to$ AdS$_5$ 
or  AdS$_5$-black brane $\to$ singularity $\to$ dS$_5$.  Although, it is known that one  of the two branches of the solution ($f_+$)  
is unstable \cite{Boulware:1985wk}  
one expects that the first two of these transitions have a holographic interpretation  
as the flow  between  two CFTs of different central charge through 
a singularity. 

(2) The  thermodynamic behaviour of charged GB black brane  is universal, \textit{i.e.}, when expressed in terms of effective mass and temperature is  
indistinguishable from that of the RN black brane.

(3)  Higher curvature terms modify the asymptotics (the AdS length) of the 5$d$ AdS-RN gravity leaving unchanged the AdS$_2\times R_3$,  extremal near-horizon geometry of the RN black brane. 
 At thermodynamic level, when expressed in terms of $M_{ADM}$ and $T_H$ a dependence  
on the normalization factor $N$ of the metric  is introduced but not 
for the extremal, near-horizon geometry AdS$_2\times R_3$.
In terms of the dual CFTs, this property  can be described as a deformation of the CFT which changes the UV behaviour  but leaves 
unchanged the IR. This  behaviour is very similar to the attractor mechanism found in supergravity theories \cite{Ferrara:1996dd,Ferrara:1995ih,Ferrara:1996um,Astefanesei:2008wz}, 
where  the  AdS$_2\times R_n$ (or AdS$_2\times S_n$) geometry is always the same irrespectively from the asymptotic values of the scalar  fields.

We have also computed the shear viscosity to entropy density ratio  for the GB charged black brane both for the non-extremal and the extremal case. 
We have found that consistently with the  geometrical and  thermodynamic picture,  universality of 
$\eta/s$ is lost in the UV but is restored in the  IR. The ratio $\eta/s$ has a non-universal temperature-dependent behaviour  for non-extremal  
black branes but attains the universal $1/4\pi$ value at extremality.
This result implies that $\eta/s$ is completely determined by the IR behaviour and is completely insensitive to the UV regime of the dual QFT.  On the one hand, this is largely expected because transport features in the hydrodynamic regime 
should be determined by IR physics. On the other hand,  it is not entirely clear if this result has a general meaning or it
is a just a consequence  of the peculiarities of the charged GB black brane (higher curvature corrections  vanish on the AdS$_2\times R_3$ background). 

Although the lesson  to be drawn from our results is that  probably it is not wise to look at the UV physics to infer about bounds  
on $\eta/s$, the question  about the possible existence  of bounds  on $\eta/s$ lower than the KSS one remains still open. We have found that $\eta/s$  is a smooth monotonic function of the temperature. Going to small temperatures,  it always flows to the 
universal value $1/4\pi$ but this value is a  \emph{minimum}  for $\lambda<0$   and \emph{maximum}
for $\lambda>0$.
Thus, the QFT dual to GB-Maxwell gravity with $\lambda<0$ gives a nice  example of  temperature-flow  of $\eta/s$ always bounded from below by $1/4\pi$. 
On the other hand, the   KSS-bound-violating flow we  obtain in the theory for   
$0<\lambda<1/4$ remains open to further investigations.

\appendix

\section{ The Black Hole case }\label{sec:BH}

This paper has been focused on the charged black brane solutions of GB gravity. 
However we would conclude with some comments on the black hole solutions of the theory, \textit{i.e.}, solutions  with $\kappa=1$ in Eq. \eqref{eq:poly}.
In the case of spherical black holes  the  discussion considerably changes. 
In fact, in 5$d$, from Eq. \eqref{eq:poly} we find that the metric  function can be written as
\begin{equation}\label{bhmetric}
 f(r)=1+\frac{r^2}{2\alpha_2}\left[1\mp\sqrt{1-4\alpha_2\left(\alpha_0-\frac{\omega_5M}{r^4}+\frac{4\pi}{3}\frac{G_5Q^2}{r^6}\right)}\right],
\end{equation}
where $\omega_5=\frac{16\pi G_5}{3\Sigma_3}$ and $\Sigma_3$ is the volume of the $3$-sphere. 
We have two branches of solutions, but similarly to the BB case, the only one admitting horizon solutions is $f_-$ with $\a_0,\a_2$ constrained by \eqref{v1}. 
The black hole mass,  
can be expressed in terms of the  horizon radius $r_+$ \cite{Cai:2003kt}
\begin{equation}
 M=\frac{r_+^4}{\omega_5}\left[\alpha_0+\frac{\alpha_2}{r_+^4}+\frac{1}{r_+^2}+\frac{4\pi G_5Q^2}{3r_+^6}\right].
\end{equation}
Due to the presence of the curvature ($\kappa=1$), now the mass depends explicitly both on the AdS radius, 
$L^2=\alpha_0^{-1}$ and on the GB coupling constant, $\alpha_2$.

The other important aspect which makes black holes different from  black branes  is that  also temperature and entropy 
depend explicitly from  $\alpha_2$ through the 
coupling with the curvature  since all the higher curvature corrections 
(like the Gauss Bonnet one) enter in the expression for the temperature trough a coupling  
with $\kappa$. As found by Cai \cite{Cai:2003kt}, for a charged 5$d$ GB black hole one gets
\begin{equation}
 T=\frac{1}{4\pi r_+(r_+^2+2\alpha_2)}\left[4\alpha_0r_+^4+2r_+^2-\frac{4\pi G_5Q^2}{3r_+^4}\right],\qquad 
 S=\frac{\Sigma_3r_+^3}{4G_5}\left(1+\frac{6\alpha_2}{r_+^2}\right).
\end{equation}
We see that since $M,T,S$ depend explicitly on the GB coupling constant $\a_2$, differently  from the black branes case,  
it is not anymore true  that the thermodynamic behaviour of the Reissner-Nordstr\"{o}m and Gauss-Bonnet black hole is the same. 
From the previous equation one can also realize  
that for the entropy, the area law  no longer holds and that it receives a correction from $\alpha_2$.\\

Let us now consider the extremal of the GB black hole. In the BB case we have found the remarkable property that 
the extremal, near-horizon  solution of the charged GB black brane is exactly the same as the RN black brane.
One can easily show that this is not the case for the extreme, near-horizon GB black hole.
In the RN case the  extremal, near-horizon, solution, which actually is an exact solution of the field equation is the  
AdS$_2\times S_3$ geometry ($S_3$ is the three sphere), 
i.e the direct product of two maximally symmetric spaces, respectively 
with negative curvature $R^{(2)}=-2/l^2$ and positive curvature $R^{(3)}= \Lambda$, 
where $l$ and $\Lambda$ can be written in terms of the 5$d$ cosmological constant and the $U(1)$ charge $Q$.

Using Eqs. \eqref{fds} one can show that  the individual   contributions of  
the AdS$_2$ and $S_3$ spaces,  to  the  two terms in Eq. \eqref{EOM} that are quadratic in the curvature tensors  vanish. 
Nevertheless there are still some cross-product contributions arising from the mixing of AdS$_2$ and $S_3$ terms. 
Splitting the 5d indices $(a,b)$ into  $\mu,\nu=0,1$ (running on AdS$_2$) and $i,j=1,2,3$ (running on $S_3$)  we find a contribution to the 
$\mu,\nu$ components of the  field equations given by $ 2\a_2\Lambda/l^2 g_{\mu\nu}$ and a contribution 
$ 4\a_2\Lambda/3l^2 g_{ij}$  to the $i j$ components of the  field equations.

We see that the  AdS$_2\times S_3$ solution of the RN field equations cannot be also  solution of the GB field equations.
Obviously, this not prevents the existence of a {\emph {different}} AdS$_2\times S_3$ solution, 
\textit{i.e}  a solution with different curvatures for AdS$_2$  and $S_3$.  
However, from the structure of the field equations and from Eqs. \eqref{fds} 
one can infer that these solutions, if existing, imply a  dependence of $l$ and/or $\Lambda$ not only  
from the  5d cosmological contant and from the black hole charge $Q$  but also from the  GB coupling constant $\a_2$.

%%%%%%%%%%%%%%%%%%%%%%%% R E F E R E N C E S
%\bibliographystyle{plain}     % bibtex style file JHEP
%\bibliographystyle{revtex4-1}
%\bibliographystyle{99}
%\bibliography{QNMbib}{}

\providecommand{\href}[2]{#2}\begingroup\raggedright
\endgroup

\end{document}